\documentclass[a4paper]{jpconf}
\usepackage{graphicx}
\begin{document}
\title{The Temperature Evolution of the Out-of-Plane Correlation Lengths of Charge-Stripe Ordered La$_ {1.725}$Sr$_{0.275}$NiO$_ 4$}

\author{P. G. Freeman$^1$, N. B. Christensen$^{2,3,4}$, D. Prabhakaran$^5$and A. T. Boothroyd$^5$ }

\ead{freeman@ill.fr}

\address{$^1$Institut Laue-Langevin, BP 156, 38042 Grenoble Cedex 9, France}

\address{$^2$Laboratory for Neutron Scattering, ETH Z\"{u}rich and Paul Scherrer Institut, CH-5232 Villigen PSI, Switzerland}

\address{$^3$Materials Research Division, Ris{\o} DTU, Technical University of Denmark, 
DK-4000 Roskilde, Denmark}

\address{$^4$Nano-Science Center, Niels Bohr Institute, University of
Copenhagen, DK-2100 Copenhagen, Denmark}

\address{$^5$Department of Physics, Oxford University, Oxford, OX1 3PU, United Kingdom}

\ead{freeman@ill.fr}

\begin{abstract}

The temperature dependence of the magnetic order of stripe-ordered La$_{1.725}$Sr$_{0.275}$NiO$_
4$ is investigated by neutron diffraction.  Upon cooling, the widths of
the magnetic Bragg peaks are observed to broaden. The degree of broadening is
found to be very different for $l =$ odd-integer and $l =$ even-integer magnetic peaks.
We argue that the observed behaviour is a result of competition between magnetic and charge order.

\end{abstract}

\section{Introduction}


Recently it has been shown both experimentally and theoretically
that charge order in La$_{2-x}$Ba$_{x}$CuO$_4$ at $x = 0.125$
decouples the Cu--O planes leading to 2D superconductivity
intimately connected to spin stripe
order\cite{2Dateighth}.
This contrasts with the stripe-ordered nickelates
La$_{2-x}$Sr$_{x}$NiO$_{4+\delta}$ (LSNO)\cite{chen-PRL-1993}, in
which the charge-stripe ordering on adjacent Ni--O planes is
correlated\cite{wochner-PRB-1998}. Inter-layer correlations in LSNO
are facilitated by a combination of a strong electron--phonon
coupling, which is also responsible for the insulating nature of
this material\cite{electron-phonon}, and Coulomb repulsions, which
cause the stripes to run parallel to one another on adjacent layers
and which control the stacking of the stripes. However, the relative importance of the
factors which control the inter-layer correlations in LSNO have not been
investigated in detail.

Charge-stripe order in LSNO occurs over a wide doping
range\cite{yoshizawa-PRB-2000} and is relatively well correlated
($\sim100$\,\AA\ in the Ni--O plane). The order is static on the timescales of
diffraction probes
\cite{chen-PRL-1993,yoshizawa-PRB-2000,neutron,neutron2,lee-PRB-2001,Kajimoto-PRB-2001,freeman-PRB-2004,giblin-PRB-2008,x-ray,HATTON}.
On cooling, the doped holes order into Ni$^{3+}$ stripes orientated at 45$^{\circ}$ to the Ni--O bonds in
the Ni--O layers. At a lower temperature, antiferromagnetic order of
the Ni$^{2+}$ spins develops with the charge stripes acting as
anti-phase domain walls for the magnetic order. Spin degrees of
freedom also exist on the Ni$^{3+}$ sites, but  coupling between
these and the magnetic order of the Ni$^{2+}$ spins, frustrates Ni$^{3+}$ spin ordering\cite{boothroyd-PRL-2003}.
The in-plane period of the charge order is given by $1/\varepsilon$ lattice units, where $\varepsilon
\approx x$. Stripes running along the $[1,-1,0]$ diagonal give rise
to charge-order diffraction peaks at positions
$(h,k,l)\pm(\varepsilon, \varepsilon,0)$ in reciprocal space, where
$h,k$ are integers  and $l$ is an odd integer. The corresponding magnetic Bragg reflections occur
at $(h+\frac{1}{2}, k+\frac{1}{2},l)\pm
(\varepsilon/2,\varepsilon/2,0)$. The tetragonal crystal structure
means that stripes running along the $[1,1,0]$ diagonal are equally
likely, and an equal population of both types of stripe domains
occurs in practice. In LSNO the charge stripes are thought to be predominately predominately Ni-centred but with some oxygen character\cite{li-PRB-2003,schu-PRL-2005}.

The temperature evolution of the in-plane stripe order has been
reported in several studies \cite{neutron2,lee-PRB-2001,Kajimoto-PRB-2001,freeman-PRB-2004,giblin-PRB-2008}, but the out-of-plane order is
much less well characterised\cite{schlappa}.
When the average in-plane periodicity $1/\varepsilon$ is
incommensurate the stripe pattern is described by the introduction
of discommensurations into a commensurate stripe order \cite{yoshizawa-PRB-2000, wochner-PRB-1998}. There are no correlations between discommensurations in adjacent
Ni--O layers \cite{wochner-PRB-1998}. In the case of $[1,-1,0]$
stripes, the basic body centred stacking of the stripes along the $c$ axis tends
to give rise to magnetic peaks with $l$ an odd integer. However,
even-integer $l$ magnetic peaks are sometimes observed as well, and
these are thought to be due to stacking faults along the $c$
direction. Evidence for this is the observation that the $l = $ even
reflections are broader in the out-of-plane direction than the $l$ =
odd reflections\cite{freeman-PRB-2004}. Hence, the correlation
lengths for the $l = $ odd and $l = $ even mangetic peaks carry
information on the coherence of the magnetic order along the $c$
direction. Correlation lengths of the spin order are larger than those of the charge order as the coherence of the order in the domains is disrupted by mainly elastic deformations\cite{yoshizawa-PRB-2000,freeman-PRB-2004,Zachar}.


Here we report on a study of the temperature dependence of the
incommensurate magnetic order in stripe-ordered
La$_{2-x}$Sr$_{x}$NiO$_{4+\delta}$ with $x = 0.275$. We find that on
warming from base temperature the magnetic Bragg reflections ${\it sharpen}$
in the out-of-plane direction above 70\,K, with the sharpening of
the $l =$ even reflections being significantly greater than that of
the $l =$ odd reflections. We show that it is necessary to take this unusual broadening into account in order to arrive at a consistent picture of the physics of La$_{2-x}$Sr$_{x}$NiO$_{4+\delta}$.
We discuss the implications of our findings for the balance between
charge and spin ordering processes in La$_ {2-x}$Sr$_ {x}$NiO$_
{4+\delta}$.

\section{Experimental details}

Single crystals of La$_{1.725}$Sr$_{0.275}$NiO$_{4+\delta}$ were
grown by the floating-zone technique\cite{prab}. The sample used
here was cut from the crystal used in our previous neutron studies
of La$_{1.725}$Sr$_{0.275}$NiO$_{4+\delta}$
\cite{freeman-PRB-2004,boothroyd-PhysicaB}, and was a rod of 8\,mm
diameter and 15\,mm length. Thermo-Gravimetric Analysis (TGA) of an
as-grown crystal determined the oxygen excess to be $\delta = 0.02\,
\pm\,0.01$. Data on the bulk magnetization of an as-grown crystal
are published elsewhere\cite{freeman-PRB-2006}. The properties of
the present crystal are consistent with those of stoichiometric LSNO
$x = 0.275$ samples studied by ourselves and
others\cite{lee-PRB-2001,Woo}.

Neutron diffraction measurements were performed on the cold
Triple-Axis Spectrometer (TAS) RITA II at the P.S.I. and the thermal
TAS IN3 at the I.L.L. On both instruments the initial and final
neutron energies were selected by  pyrolytic
graphite  vertically-focusing monochromators,  a vertically-focusing 
analyzer on IN3 and a flat  analyzer on RITA II. Higher-order harmonics were
suppressed after the sample and before the analyzer by use of a
liquid nitrogen-cooled Be filter on RITA II and a PG filter on IN3.
On IN3 the horizontal divergence of the neutrons was constrained by
30\'{} and 20\'{} collimators placed in the incident and scattered beams respectively. The
neutron energies employed on RITA II and IN3 were $E = 5$\,meV and
$14.7$\,meV, respectively.
A standard I.L.L. orange cryostat was used as the sample environment
on IN3, and a $15$\,T Oxford Instrument cryomagnet  was used on RITA
II. The sample was mounted in a cryomagnet for a study of the effect
of a magnetic field on the magnetic order of
La$_{1.725}$Sr$_{0.275}$NiO$_{4+\delta}$, which will be reported
elsewhere. The sample was oriented on both instruments so that $(h,
h, l)$ positions in reciprocal space could be accessed. In this work
we refer to the tetragonal unit cell of LSNO, which has unit cell
parameters $a \approx$ 3.8 \AA\ and $c \approx$ 12.7 \AA. The
crystal was pre-aligned for the measurements on the neutron
diffractometer Morpheus at P.S.I. and the neutron Laue
diffractometer Orient Express at the I.L.L.

\section{Results}

\begin{figure}[!h]
\begin{center}
\includegraphics[clip=, width = 16cm]{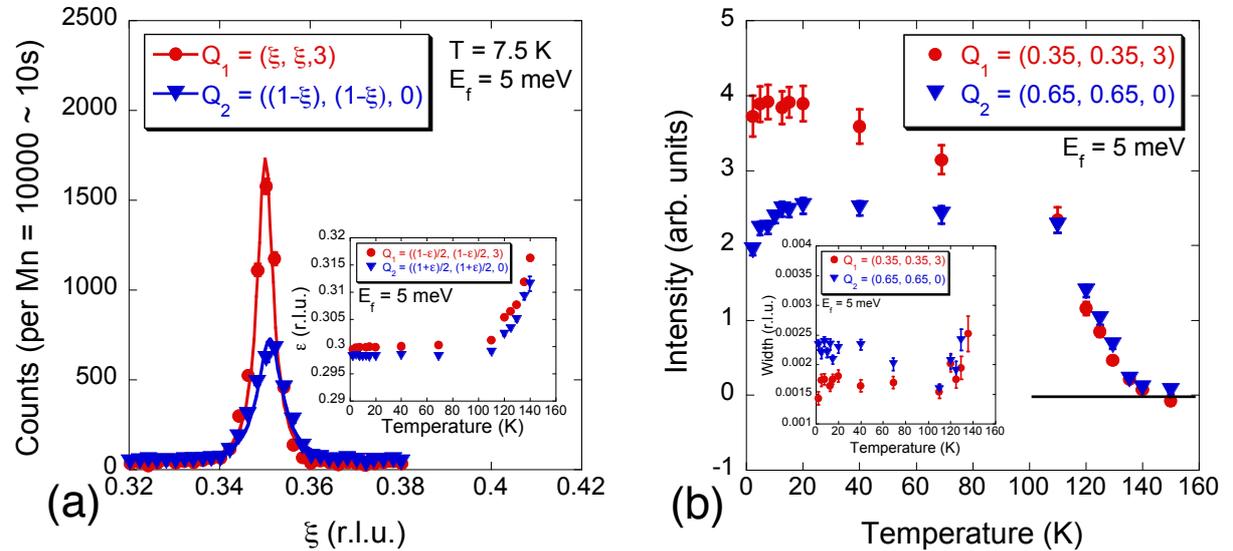}
\caption[(hh0) scans of two magnetic Bragg reflections]{(colour
online) (a) Neutron diffraction from two magnetic Bragg reflections
of stripe-ordered La$_{1.725}$Sr$_{0.275}$NiO$_{4+\delta}$ measured
parallel to $(h,h,0)$. The lines are fits to a Gaussian function
on a flat background. Inset: Temperature variation of the
incommensurability $\varepsilon$ obtained from the magnetic Bragg
reflections ${\bf Q}_1$ and
${\bf Q}_2$.
(b) Temperature dependence of the integrated intensities (Gaussian width multiplied by Gaussian amplitude) of the two magnetic Bragg
reflections obtained from scans parallel to $(h,h,0)$.  Inset: Temperature variation of the resolution corrected Gaussian widths of the magnetic Bragg reflections measured in scans parallel to $(h,h,0)$. Along $(h,h,0)$
1\,r.l.u. = 2.32 \AA$^{-1}$. 
} \label{fig:hhdirection}
\end{center} \end{figure}

In figure \ref{fig:hhdirection}(a) we show magnetic Bragg
reflections from La$_{1.725}$Sr$_{0.275}$NiO$_{4}$, measured along  $(h,\ h,\ 0)$  at two
positions in reciprocal space, ${\bf Q}_1 = (0.35, 0.35,3)$ and
${\bf Q}_2 = (0.65, 0.65,0)$. The data are consistent
with the general position  $(h+\frac{1}{2}, k+\frac{1}{2},l)\pm
(\varepsilon/2,\varepsilon/2,0)$ with $\varepsilon = 0.299 \pm
0.004$. The slight difference in the centering of the two peaks is
due to imperfect alignment of the crystals ($\sim 1^{\circ}$) in the
cryomagnet which could not be tilted.  The ordered moments of this
compound have previously been found to lie in the $ab$
plane\cite{freeman-PRB-2004}. Since neutrons scatter from spin
components perpendicular to the scattering vector, ${\bf Q}_1$ is to
a good approximation sensitive to the total in-plane moment, while
${\bf Q}_2$ is only sensitive to the spin component parallel to the
stripe direction $([1,-1, 0]$. We have fitted the peaks to
Gaussian functions.
${\bf Q}_1 = (0.35, 0.35,3)$ and  ${\bf Q}_2 = (0.65, 0.65, 0)$  peaks are slightly broadened compared to the estimated experimental resolution of 0.00220$\,\pm\,$0.00002\,r.l.u.  and 0.00294$\,\pm\,$0.00005\,r.l.u.respectively, for scans along the $(h,\ h,\ 0)$ direction, based on a fit to structural Bragg
reflections.

We plot in  the inset of Fig. \ref{fig:hhdirection}(a) the temperature dependence
of $\varepsilon$ of the magnetic reflections plotted
in Fig. \ref{fig:hhdirection}(a). Consistent with previous
observations, on warming 
$\varepsilon$ remains constant up to 110\,K, then tends to a value of $\varepsilon=1/3$ as the
magnetic order melts at $T_{\rm SO} = 140 \pm 5$\,K \cite{Kajimoto-PRB-2001}. 
The small difference between the values of $\varepsilon$ measured at ${\bf Q}_1$ and
${\bf Q}_2$ is partly caused by the sample tilt mentioned above, but
the increase in the discrepancy at $T>100$\,K is probably because
the sample temperature was not perfectly in equilibrium during the
two measurements.

\begin{figure}[!ht]
\begin{center}
\includegraphics[clip=, width = 16cm]{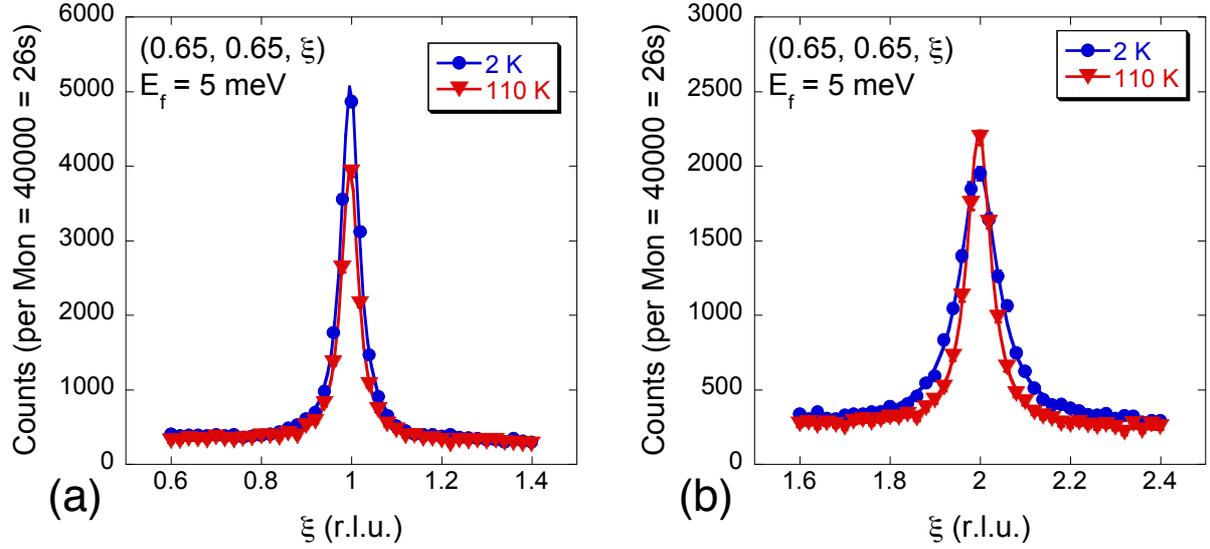}
\caption[(00l) scans of two magnetic Bragg reflections at two
temperatures]{(colour online) Scans parallel to $(0,0,l)$ of the
magnetic Bragg reflections (a) $(065,0.65, 1)$ and (b)
$(0.65,0.65,2)$ of La$_{1.725}$Sr$_{0.275}$NiO$_{4+\delta}$ at 2\,K
and 110\,K. The lines are the results of fits to a Lorentzian
function with a flat background. In (b) the magnetic reflection is
clearly observed to sharpen between 2\,K and 110\,K.
}
\label{fig:twoTs}
\end{center} \end{figure}

\begin{figure}[!h]
\begin{center}
\includegraphics[clip=, width = 16cm]{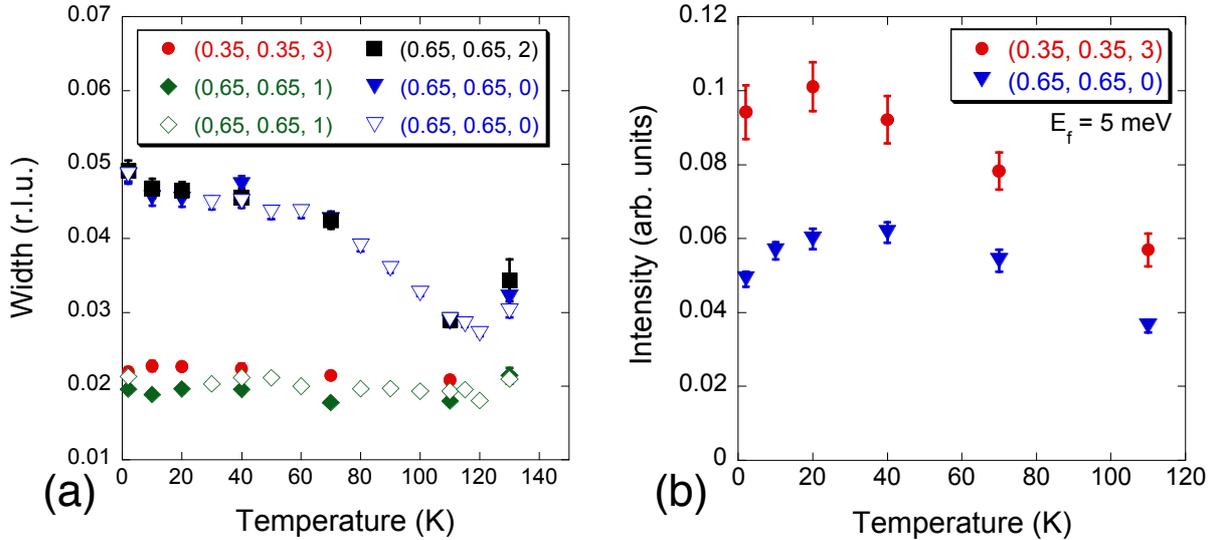}
\caption[Temperature variation in l direction.]{(colour online) (a) Temperature variation of the resolution corrected  Lorentzian widths along $(0, 0, l)$ of several magnetic Bragg
reflections from La$_{1.725}$Sr$_{0.275}$NiO$_{4}$. Solid symbols
are the results from RITA II and open symbols are results from IN3.
Between 70\,K and 110\,K the width of all of the Bragg
reflections is observed to sharpen, but the sharpening is much more
significant for the $l =$ even reflections. Along $(0, 0,l)$, 1
r.l.u. = 0.497\,\AA $^{-1}$. (b) Temperature dependence of the
integrated intensity of ${\bf Q}_1 = (0.35, 0.35,3)$ and ${\bf Q}_2
= (0.65, 0.65,0)$, obtained from the product of the Lorentzian
widths in the $(h, h, 0)$ and $(0, 0,l)$ directions and the
Lorentzian amplitude. Between 40\,K and 110\,K the ratio of the
intensities of the two reflections remains constant to a good
approximation. } \label{fig:tvariationldep}
\end{center} \end{figure}


The temperature variation of the resolution corrected widths of the magnetic Bragg
reflections measured in  
 $(h,h,0)$ scans is shown in the inset of figure \ref{fig:hhdirection}(b).  Instrument resolution was estimated using the widths of nearby  structural Bragg reflections. On warming the width of the ${\bf
Q}_1$ magnetic Bragg reflection remains approximately constant up to
110\,K before broadening as the spin ordering temperature is
approached.  The ${\bf Q}_2$ magnetic Bragg reflection sharpens by
$27 \pm 5 $\,\% between 40\,K and 110\,K above which it broadens as
the spin ordering temperature is approached.  For the ${\bf Q}_2$  reflection, between 40\,K and 110\,K the peak slightly  sharpens by $11 \pm 6\,\%$.




In  Fig.\ \ref{fig:hhdirection}(b) we show the integrated intensities of the ${\bf Q}_1$ and ${\bf Q}_2$
reflections from scans parallel to $(h,h,0)$. No correction has been made for the small
difference between the Ni$^{2+}$ magnetic form factor at the ${\bf
Q}_1$ and ${\bf Q}_2$ wavevectors.  On warming from 2\,K the intensity of ${\bf
Q}_2$ increases up to 20\,K due to a spin re-orientation found previously\cite{freeman-PRB-2004}. At higher temperatures the ratio   of the intensities of the ${\bf Q}_1$ and ${\bf Q}_2$ peaks is expected to remain constant, but it actually decreases between 40\,K and $T_{\rm SO} = 140 \pm 5$\,K. 

Next we describe the temperature dependence of different magnetic
Bragg reflections in scans parallel to $(0,0,l)$. In Figs.\
\ref{fig:twoTs}(a) and (b) we show $(0,0,l)$ scans through
the magnetic Bragg reflections $(0.65, 0.65, 1)$ and $(0.65, 0.65,
2)$, respectively, at both 2\,K and 110\,K. The solid lines indicate
a fit to a Lorentzian function on a sloping background. The $(0.65,
0.65, 1)$ reflection in Fig. \ref{fig:twoTs}(a) loses amplitude
between 2\,K and 110\,K with no significant change in width, whereas
the $(0.65,0.65,2)$ reflection shown in Fig. \ref{fig:twoTs}(b) is
sharper at 110\,K compared with 2\,K and the amplitude of the peak
slightly increases.

Figure \ref{fig:tvariationldep}(a) shows the temperature variation
of the resolution corrected $l$ widths for several different magnetic Bragg reflections.
For the $l = $ even reflections, the widths of the
peaks decrease very slightly on warming up to 70\,K, then between
70\,K and 120\,K the peaks sharpen dramatically before broadening as
the temperature goes above 120\,K. The $l =$ odd reflections
decrease very slightly in width between 40\,K and 120\,K then
broaden at higher temperatures. For comparison, the sharpening of
the $l =$ even peaks between 10\,K and 120\,K is $38\pm 2$\,\%,
whereas for the $l =$ odd reflections it is only $7\pm 3$\,\%.

In Fig.\ \ref{fig:tvariationldep}(b) we show the integrated
intensities of the $(0.35,0.35,3)$ and $(0.65,0.65,0)$ magnetic
Bragg reflections as a function of temperature,
 obtained from the product of the Gaussian and Lorentzian widths in
the $(h, h, 0)$ and $(0, 0, l)$ directions respectively, and the Gaussian
amplitude.  Between $40$\,K and $110$\,K the ratio of the integrated
intensities of two reflections, which have $l =$ odd and $l =$ even
respectively, is to a good approximation constant. This determines
that the greater sharpening in the $l$ direction of the $l =$ even
reflections relative to the $l =$ odd reflections is responsible for
the different temperature dependences of the integrated intensities
in scans parallel to $(h, h, 0)$ shown in Fig.\
\ref{fig:hhdirection}(b). The widths in the $(h,-h,0)$ direction
have not been included in the analysis,  as the instrument resolution out of the scattering plane is large enough to integrate over all the intensity in the $(h,-h,0)$ direction, irrespective any broadening.


\section{Discusion and conclusions}

The central results of this study of stripe-ordered
La$_{1.725}$Sr$_{0.275}$NiO$_{4}$ are, (i)  as magnetic order develops the magnetic Bragg peaks
broaden along the $c$-axis, and (ii) the temperature-dependent broadening of the $l =$
even-integer reflections is 5 times greater (38\,\% compared to 7\,\%) than that of the $l =$ odd-integer reflections, see Fig.\ \ref{fig:twoTs} and
Fig.\ \ref{fig:tvariationldep}(a).
The reduction in out-of-plane
correlation length on cooling is unrelated to the
previously-observed spin re-orientation\cite{freeman-PRB-2004}.



It is believed that the ordering processes in LSNO are driven by charge order, because charge ordering occurs at a higher temperature
than spin ordering \cite{yoshizawa-PRB-2000,Kajimoto-PRB-2001}.
Previous studies show that on cooling the spin--charge stripe
periodicity changes from that preferred by the charge order, to a
periodicity that is a compromise between that preferred by the spin
order and that preferred by the charge order
\cite{Kajimoto-PRB-2001}\cite{footnote}.
With reference to the new observations reported here, this suggests
that in La$_{1.725}$Sr$_{0.275}$NiO$_{4}$ the charge order favours
long-range order along the $c$ axis whereas the combination of charge order and fully-developed magnetic
order tends to disrupt the coupling between the layers, perhaps due to frustration effects.

Let us now consider why the $l =$ even magnetic reflections broaden
in the out-of-plane direction to a greater degree on cooling than do
the $l = $ odd reflections.  The difference lies in the origin of
these two reflections \cite{wochner-PRB-1998,freeman-PRB-2004}. The
nature of the spin order in the out-of-plane direction is primarily controlled by
minimization of Coulomb repulsions between the charge stripes. For
commensurate $x = 1/3$ stripes, the stacking of the spin--charge
order along the $c$ axis results in dominant $l =$ odd magnetic Bragg
reflections \cite{lee-PRB-2001}. Incommensurate stripe order, such as that at
$x=0.275$, cannot adopt an ideal $x = 1/3$ stacking and exhibits $l =$ even magnetic reflections in addition to $l =$ odd
reflections. The fact that the $l=$ even and $l=$ odd peaks have different $l$ widths
suggests the presence of more than one stacking sequence, with
different sequences having different correlation lengths along the $c$ axis.
Upon cooling, the development of magnetic order must introduce stacking faults
along the $c$ axis in such a way that the majority $x = 1/3$ -type stacking
is relatively unaffected while the minority stacking which has the largest
influence on the $l=$ even reflections becomes less well correlated. The cause of the  smaller broadening of the $l=$ even magnetic Bragg reflections in the  $(h,h,0)$ direction could be related
to the large reduction of coherence in the magnetic order along the $c$ axis.

Our results are complementary to data recently reported by Schlappa {\it et al.}, who studied the charge and spin ordering in
incommensurate stripe-ordered La$_{1.8}$Sr$_{0.2}$NiO$_{4+\delta}$
by resonant x-ray scattering (RXS) techniques and neutron diffraction \cite{schlappa}. The main focus of their study was on the temperature
dependence of the intensity of the charge and magnetic Bragg peaks measured by different diffraction probes,
but they also found an increase in the widths of both
the charge and magnetic Bragg peaks with decreasing temperature. As pointed out by Schlappa {\it et al.},
care needs to be taken when comparing magnetic and charge Bragg peaks measured by neutron and x-ray scattering since
the x-ray measurement integrates over all the fluctuation spectrum in addition to the static component of the order.
Nevertheless, the similar temperature dependence of the widths of the magnetic reflections measured by RXS  and the neutron diffraction in this study,  suggests that the broadening of the charge peaks with decreasing temperature observed by x-rays is likely a property of
the static order. This provides further evidence that the increased disorder in the $c$-axis stacking at low temperatures is a result of coupling between the magnetic and charge order.



As mentioned earlier, charge order in the 1/8-doped cuprate La$_ {2-x}$Ba$_{x}$CuO$_4$
tends to decouple the Cu-O planes, leading to
2D superconductivity and quasi-2D magnetic ordering. In stripe-ordered
La$_{1.725}$Sr$_{0.275}$NiO$_{4+\delta}$ charge order couples the Ni--O planes
favouring quasi-3D order. As magnetic order develops, however, charge and magnetic interactions
along the $c$ axis compete and tend to reduce the out-of-plane correlation lengths, making
the spin--charge ordering more 2D.
Therefore, LSNO and the 1/8-doped cuprate behave in a qualitatively similar way as far as the
inter-layer coupling of the magnetic order is concerned, but differ in that the charge order on adjacent layers in LSNO is much more strongly coupled than it is in the cuprate.


\section{Acknowledgements}

The work was performed mainly at the Swiss Spallation Neutron Source SINQ, Paul
Scherrer Institut (PSI), Villigen, Switzerland. This research project
has been supported by the European Commission under the 7th Framework Programme
through the 'Research Infrastructures' action of the 'Capacities' Programme,
Contract No: CP-CSA\_INFRA-2008-1.1.1 Number 226507-NMI3.  N. B. Christensen
would also like to acknowledge support from the Swiss NSF grant 200020-105175. Crystal growth
at Oxford was supported by the Engineering and
Physical Sciences Research Council of Great Britain.

\end{document}